\documentclass[12pt,twoside]{article}

\usepackage{moriond,epsfig}

\usepackage{times}
\usepackage{cite}

\usepackage{cite}   

\newcommand{\Bbar}{\,\overline{\!B}}
\newcommand{\bb}{\ensuremath{B^0\!-\!\Bbar{}^0\,}}
\newcommand{\bbq}{\ensuremath{B_q\!-\!\Bbar{}_q\,}}
\newcommand{\bbd}{\ensuremath{B_d\!-\!\Bbar{}_d\,}}
\newcommand{\bbs}{\ensuremath{B_s\!-\!\Bbar{}_s\,}}
\newcommand{\kkm}{$K^0\!-\ov{\!K}{}^0\,$ mixing}
\newcommand{\bra}[1]{\ensuremath{\langle #1 |}}
\newcommand{\ket}[1]{\ensuremath{| #1 \rangle }}

\newcommand{\gqtf}{\ensuremath{\Gamma (B_q(t) \rightarrow f )}}
\newcommand{\gqbtf}{\ensuremath{\Gamma (\Bbar{}_q(t) \rightarrow f )}}
\newcommand{\gqtfb}{\ensuremath{\Gamma (B_q(t) \rightarrow \ov{f} )}}
\newcommand{\gqbtfb}{\ensuremath{\Gamma (\Bbar{}_q(t) \rightarrow \ov{f} )}}

\newcommand{\gdbtf}{\ensuremath{\Gamma (\Bbar{}_d(t) \rightarrow f )}}
\newcommand{\gdtfb}{\ensuremath{\Gamma (B_d(t) \rightarrow \ov{f} )}}

\newcommand{\guntf}{\ensuremath{\Gamma  [f,t] }}
\newcommand{\guntfb}{\ensuremath{\Gamma  [\ov{f},t] }}

\newcommand{\dm}{\ensuremath{\Delta M}}
\newcommand{\dg}{\ensuremath{\Delta \Gamma}}
\newcommand{\lqcd}{\Lambda_{\rm QCD}}

\newcommand{\ds}{\displaystyle}
\newcommand{\lt}{\left}
\newcommand{\rt}{\right}
\newcommand{\no}{\nonumber}
\newcommand{\nn}{\nonumber \\}

\newcommand{\ov}[1]{\overline{#1}}
\newcommand{\eq}[1]{Eq.~(\ref{#1})}
\newcommand{\imag}{\mathrm{Im}\,}
\newcommand{\real}{\mathrm{Re}\,}

\newcommand{\adi}{A^{\rm dir}}
\newcommand{\ami}{A^{\rm mix}}
\newcommand{\adg}{A^{\rm \Delta\Gamma}}

\newcommand{\BsorBsbar}{\raisebox{7.7pt}{$\scriptscriptstyle(\hspace*{8.5pt})$}
  \hspace*{-10.7pt}\!\Bbar_{s}}

\newcommand{\NorNbar}{\raisebox{8.2pt}{$\scriptscriptstyle(\hspace*{9.5pt})$}
  \hspace*{-11.7pt}\ov{\!N}}

\begin{document}
\thispagestyle{empty}
\title{hep-ph/0406300 \hfill   FERMILAB-Conf-04/094-T\\[1cm]
\Large CP asymmetry in flavour-specific 
B decays~\footnote{ 
Talk presented at the Moriond conference on 
\emph{Electroweak Interactions and Unified Theories, 2004}.
}}

\author{Ulrich Nierste}

\address{Fermi National Accelerator Laboratory, Batavia,
        IL 60510-500, USA.}

\maketitle\abstracts{
I first discuss the phenomenology of $a^q_{\rm fs}$ ($q=d,s$), which is the
CP asymmetry in flavour-specific $B_q$ decays such as $B_d \to X \ell^+
\ov{\nu}_\ell$ or $B_s \to D_s^- \pi^+$.  $a^q_{\rm fs}$ can be obtained
from the time evolution of \emph{any}\ untagged $B_q$ decay.  Then I
present recently calculated next-to-leading-order QCD corrections to
$a_{\rm fs}^q$, which reduce the renormalisation scheme uncertainties
significantly. For the Standard Model we predict $a^d_{\rm fs}=-(5.0
\pm 1.1)\times 10^{-4}$ and $a^s_{\rm fs}= (2.1 \pm 0.4)\times
10^{-5}$. As a by-product we determine the ratio of the width
difference in the $B_d$ system and the average $B_d$ width to
$\dg_d/\Gamma_d = (3.0\pm 1.2)\times 10^{-3}$ at next-to-leading order
in QCD.
}

\section{Preliminaries}
The time evolution of the \bbd\ system is
determined by a Schr\"odinger equation:
\begin{eqnarray}
 i \frac{d}{d\, t} \left( \!
\begin{array}{c}
\ds \ket{B_d(t)} \\[1mm]
\ds \ket{\ov{B}_d(t)}
\end{array}\! \right) &  = & 
\left( M^d - i\, \frac{\Gamma^d}{2}  \right)
\left(\!
\begin{array}{c}
\ds \ket{B_d(t)} \\[1mm]
\ds \ket{\ov{B}_d(t)}
\end{array}\!\right), \label{sch}
\end{eqnarray}
which involves two Hermitian $2\times 2$ matrices, the mass matrix
$M^d$ and the decay matrix $\Gamma^d$. Here $B_d(t)$ and
$\Bbar{}_d(t)$ denote mesons which are tagged as a $B_d$ and $\Bbar_d$
at time $t=0$, respectively. 
By diagonalising  $M^d - i\Gamma^d/2$ one obtains the mass eigenstates:
\begin{eqnarray}
\mbox{Lighter eigenstate:} \quad \ket{B_{d,L}} & =&
        p \ket{B_d^0} + q \ket{\ov{B}{}_d^0} .  \nn
\mbox{Heavier eigenstate:} \quad \ket{B_{d,H}} & =&
        p \ket{B_d^0} - q \ket{\ov{B}{}_d^0}
\quad \qquad\mbox{with } \lt|p\rt|^2+\lt|q\rt|^2 = 1. \label{eig}
\end{eqnarray}
We discuss the mixing formalism for $B_d$ mesons, the corresponding
quantities for \bbs\ mixing are obtained by the replacement $d\to
s$. The coefficients $q$ and $p$ in \eq{eig} are also different for
the $B_d$ and $B_s$ systems.  The \bbd\ oscillations in \eq{sch}
involve the three physical quantities $|M_{12}^d|$, $|\Gamma_{12}^d|$
and $\phi_d=\arg(-M_{12}^d/\Gamma_{12}^d)$ (see e.g.\ \cite{run2}).
The mass and width differences between $B_{d,L}$ and $B_{d,H}$ are
related to them as
\begin{eqnarray}
\dm_d &=& M^d_H -M^d_L \; = \; 2\, |M_{12}^d|,
\qquad \dg_d \; =\; \Gamma^d_L-\Gamma^d_H \; =\;
        2\, |\Gamma_{12}^d| \cos \phi_d, \label{dmdg}
\end{eqnarray}
where $M^d_L,\Gamma^d_L$ and $M^d_H,\Gamma^d_H$ denote the masses
and widths of $B_{d,L}$ and $B_{d,H}$, respectively.

The third quantity to determine the mixing problem in \eq{sch} is 
\begin{eqnarray}
a^d_{\rm fs}
     &=& 
    \imag \frac{\Gamma_{12}^d}{M_{12}^d}
    \; = \; \frac{\dg_d}{\dm_d} \tan \phi_d
 . \label{defafs}
\end{eqnarray}
$a_{\rm fs}^d$ is the CP asymmetry in \emph{flavour-specific} $B_d\to
f$ decays, which means that the decays $\Bbar_d \to f$ and $B_d \to
\ov{f}$ (with $\ov{f}$ denoting the CP-conjugate final state) are
forbidden \cite{hw}. Next we consider flavour-specific decays in which
the decay amplitudes $A_f=\bra{f} B_d \rangle$ and
$\ov{A}_{\ov{f}}=\bra{\ov{f}} \Bbar_d \rangle$ in addition satisfy
\begin{eqnarray}
|A_f| &=& |\ov{A}_{\ov{f}}|. \label{nod}
\end{eqnarray}
\eq{nod} means that there is no direct CP violation in $B_d \to
f$. Then $a^d_{\rm fs}$ is given by 
\begin{eqnarray}
a^d_{\rm fs}
     &=& \frac{\gdbtf - \gdtfb}{\gdbtf + \gdtfb}
 . \label{afst}  
\end{eqnarray}
Note that the oscillatory terms cancel between numerator and denominator.
The standard way to access
$a_{\rm fs}^d$ uses $B_d \to X \ell^+ \ov{\nu_\ell}$ decays, which
justifies the name \emph{semileptonic CP asymmetry} for $a_{\rm
fs}^d$.  In the $B_s$ system one can also use $B_s \to D_s^- \pi^+$ to
measure $a_{\rm fs}^s$. Yet, for example, \eq{afst} does not apply to
the flavour-specific decays $B_d \to K^+ \pi^-$ or $B_s \to K^-
\pi^+$, which do not obey \eq{nod}.

$a_{\rm fs}^d$ measures \emph{CP violation in mixing}. Other commonly used 
notations involve the quantities $|q/p|$ or $\epsilon_B$; they are
related to $a_{\rm fs}^d$ as
\begin{eqnarray}
1-\lt| \frac{q}{p} \rt| &=& \frac{a_{\rm fs}^d}{2}, \qquad \qquad
\frac{\real \epsilon_B}{1+|\epsilon_B|^2} \; = \; \frac{a_{\rm
fs}^d}{4} . \label{eb}
\end{eqnarray}
Here $\epsilon_B= (1+q/p)/(1-q/p)$ is the analogue of the quantity 
$\ov{\epsilon}_K$ in \kkm. Unlike $a_{\rm fs}^d$ it depends on phase 
conventions and should not be used. In \eq{eb} and future equations we
neglect terms of order $(a_{\rm fs}^q)^2$.

$a_{\rm fs}^d$ is small for two reasons: First
$|\Gamma_{12}^d/M_{12}^d|=O(m_b^2/M_W^2)$ suppresses $a_{\rm fs}^d$ to
the percent level. Second there is a GIM suppression factor
$m_c^2/m_b^2$ reducing $a_{\rm fs}^d$ by another order of
magnitude. Generic new physics contributions to $\arg M_{12}^d$ (e.g.\
from squark-gluino loops in supersymmetric theories) will lift this
GIM suppression. $a_{\rm fs}^s$ is further suppressed by two powers of
the Wolfenstein parameter $\lambda\simeq 0.22$. Therefore $a_{\rm
fs}^d$ and $a_{\rm fs}^s$ are very sensitive to new CP phases
\cite{run2,llnp}, which can enhance $|a_{\rm fs}^d|$ and $|a_{\rm fs}^s|$
to 0.01. $|a_{\rm fs}^d|$ can be further enhanced by new contributions
to $\Gamma_{12}^d$, which is doubly Cabibbo-suppressed in the Standard
Model.

The experimental world average for $a_{\rm fs}^d$ is \cite{s}
\begin{eqnarray}
a_{\rm fs}^d &=& 0.002\pm0.013 . \no
\end{eqnarray}

\boldmath
\section{Measurement of $a_{\rm fs}^q$}
\unboldmath
\subsection{Flavour-specific decays}
We first discuss the flavour-specific decays without direct CP
violation in the Standard Model.  First note that the ``right-sign''
asymmetry vanishes:
\begin{eqnarray}
    \gqtf - \gqbtfb & = & 0 . \label{unm}
\end{eqnarray}
Since we are hunting possible new physics in a tiny quantity, we
should be concerned whether \eq{nod} still holds in the presence of
new physics.\footnote{Direct CP violation requires the presence of a
CP-conserving phase. In the case of $B_d\to D^- \ell^+\nu_{\ell}$ this
phase comes from photon exchange and is small. Also somewhat contrived
scenarios of new physics are needed to get a sizeable CP-violating
phase in a semileptonic decay.  Thus here one needs to worry about $
|A_f| \neq |\ov{A}_{\ov{f}}|$ only, once $a_{\rm fs}^d$ is probed at
the permille level.}  Further no experiment is exactly
charge-symmetric, and the efficiencies for $\Bbar \to \ov{f}$ and $B
\to f$ may differ by a factor of $1+\delta_c$. One can use the
``right-sign'' asymmetry in \eq{unm} to calibrate for both effects: In
the presence of a charge asymmetry $\delta_c$ one will measure
\begin{eqnarray}
a^{q,\delta_c}_{\rm right} &\equiv&
\frac{\gqtf - (1+\delta_c)\gqbtfb}{\gqtf + (1+\delta_c)\gqbtfb} \;=\; 
\frac{|A_f|^2-|\ov{A}_{\ov{f}}|^2}{|A_f|^2+|\ov{A}_{\ov{f}}|^2} 
-\frac{\delta_c}{2} . \label{unmc}
\end{eqnarray}
Instead of the desired CP asymmetry in \eq{afst} one will find
\begin{eqnarray}
a^{q,\delta_c}_{\rm fs}
     &=& \frac{\gdbtf -  (1+\delta_c)\gdtfb}{\gdbtf +  (1+\delta_c)\gdtfb}
 \; = \; a^q_{\rm fs} +a^{q,\delta_c}_{\rm right} . \label{afstc}  
\end{eqnarray}
Thus $\delta_c$ and the direct CP asymmetry 
$(|A_f|^2-|\ov{A}_{\ov{f}}|^2)/(|A_f|^2+|\ov{A}_{\ov{f}}|^2)$ enter 
\eq{unmc} and \eq{afstc} in the same combination and  
$a^q_{\rm fs}$ can be determined. Above we have kept only terms to
first order in the small quantities 
$1-|\ov{A}_{\ov{f}}|^2/|A_f|^2$, $\delta_c$ and $a^q_{\rm fs}$.

It is well-known that the
measurement of $a_{\rm fs}^q$ requires neither tagging nor the
resolution of the \bbq\ oscillations \cite{hw}. 
Since the right-sign asymmetry in \eq{unm} vanishes, 
the information on $a_{\rm fs}^q$ from \eq{afst} persists in
the untagged decay rate
\begin{eqnarray}
\guntf &=& \gqtf + \gqbtf      .   \label{guntf}
\end{eqnarray}
At a hadron collider one also cannot rule out a production asymmetry
$\delta_p = N_{\Bbar_q}/N_{B_q} -1$ between the numbers
$N_{\Bbar_q}$ and $N_{B_q}$ of $\Bbar_q$'s and $B_q$'s. An untagged 
measurement will give
\begin{eqnarray}
a_{\rm fs, unt}^{q,\delta_c}(t) &=&
   \frac{\guntf - (1+\delta_c) \guntfb}{\guntf + (1+\delta_c)\guntfb}
\;= \; a^{q,\delta_c}_{\rm right} + 
        \frac{a_{\rm fs}^q}{2} - \frac{a_{\rm fs}^q+\delta_p}{2} \,
        \frac{\cos (\dm_q\, t)}{\cosh (\dg_q t/2) }
        . \,  \label{fsun}
\end{eqnarray}
The use of the larger untagged data sample to determine $a_{\rm fs}^d$
seems to be advantageous at the $\Upsilon(4S)$ B factories, where
$\delta_p=0$. Then the time evolution in \eq{fsun} contains enough
information to separate $a_{\rm fs}^d$ from 
$a^{d,\delta_c}_{\rm right}= a_{\rm fs, unt}^{d,\delta_c}(t=0)$.

Eqs.~(\ref{afst}),(\ref{unmc}) and (\ref{afstc}) still hold, when the
time-dependent rates are integrated over $t$. The time-integrated
untagged CP asymmetry reads (for $ |A_f| = |\ov{A}_{\ov{f}}|$,
$\delta_c=\delta_p=0$):
\begin{eqnarray}
A_{\rm fs,unt}^q &\equiv & 
\frac{\int_0^\infty dt [ \guntf - \guntfb  ]}{
        \int_0^\infty dt [ \guntf + \guntfb] } 
\; =\; \frac{a_{\rm fs}^q}{2} \, \frac{x_q^2+y_q^2}{x_q^2+1} ,
\label{aunt}
\end{eqnarray}
where $x_q=\dm_q/\Gamma_q$,  $y_q=\dg_q/(2\Gamma_q)$ and $\Gamma_q$ is
the average decay width in the $B_q$ system. In particular a
measurement of $a_{\rm fs}^s$ does not require to resolve the rapid 
\bbs\ oscillations. In $\Upsilon(4S)$ B factories a common method to 
constrain $a_{\rm fs}^d$ is to compare the number $N_{++}$ of decays  
$(B_d(t),\Bbar_d(t))\to (f,f)$ with the number $N_{--}$ of decays 
to $(\ov{f},\ov{f})$, typically for $f=X\ell^+\nu_{\ell}$. Then
one finds $a_{\rm fs}^d=(N_{++}-N_{--})/(N_{++}+N_{--})$. 

We next exemplify the measurement of $a_{\rm fs}^s$ from 
time-integrated tagged $B_s \to f$ decays, having
$f=X\ell^+\nu_{\ell}$ in mind. This approach should be feasible  
at the Fermilab Tevatron. We allow the detector to be
charge-asymmetric ($\delta_c\neq 0$) and also relax \eq{nod} 
to $|A_f| \approx |\ov{A}_{\ov{f}}|$. Let $N_f$ denote the total number of
observed decays of meson tagged as $B_s$ at time $t=0$ into the final
state $f$. Further $\ov{N}_{f}$ denotes the analogous number for a
meson initially tagged as a $\Bbar_s$. The corresponding quantities
for the decays $B_s(t)\to\ov{f}$ and $\Bbar_s(t)\to\ov{f}$ are 
$N_{\ov{f}}$ and $\ov{N}_{\ov{f}}$. One has 
\begin{eqnarray}
\NorNbar_f \; \propto\; \int_0^\infty \!\! dt \, \Gamma (\BsorBsbar (t) \to f), 
\qquad\qquad
\NorNbar_{\ov{f}} \; \propto\; (1+\delta_c) 
\int_0^\infty \!\! dt \, \Gamma (\BsorBsbar (t) \to \ov{f}) \no
\end{eqnarray}
with the same constant of proportionality. The four asymmetries 
\begin{eqnarray}
\frac{N_f-\ov{N}_{\ov{f}}}{N_f+\ov{N}_{\ov{f}}} \; = \; 
a^{s,\delta_c}_{\rm right} \, ,
&\qquad& 
\frac{\ov{N}_f - N_{\ov{f} }}{\ov{N}_f + N_{\ov{f} }}
\; = \; 
a^{s,\delta_c}_{\rm right} + a_{\rm fs}^s \, ,\nn 
\frac{N_f-\ov{N}_f}{N_f+\ov{N}_f} 
\; =\; \frac{1-y_s^2}{1+x_s^2} - \frac{a_{\rm fs}^s}{2} \, , 
&\qquad& 
\frac{\ov{N}_{\ov{f}} - N_{\ov{f}}}{\ov{N}_{\ov{f}} + N_{\ov{f}}}
\; =\; \frac{1-y_s^2}{1+x_s^2} + \frac{a_{\rm fs}^s}{2} 
\label{nafs}
\end{eqnarray}
then allow to determine $a_{\rm fs}^s$ and $(1-y_s^2)/(1+x_s^2)$.
In the second line of \eq{nafs} terms of order $a_{\rm fs}^s/x_s^2$ 
have been neglected. (Of course the last asymmetry in \eq{nafs} is redundant.)

\subsection{Any decay}
Since $q/p$ enters the time evolution of \emph{any}\ neutral $B_q\to f$
decay, we can use any such decay to determine $a_{\rm fs}^q$. The 
time dependent decay rates involve
\begin{eqnarray}
\lambda_f &=& \frac{q}{p} \frac{\bra{f} \Bbar_q \rangle}{\bra{f} B_q \rangle}
. \no
\end{eqnarray}
In Eq.~(1.73)-(1.77) of \cite{run2} 
\gqtf, \gqbtf, \gqtfb and \gqbtfb\ can be found for the most general
case, including a non-zero $\dg_q$. For the untagged rate one easily finds
\begin{eqnarray}
\guntf &\propto & e^{-\Gamma_q t} \lt\{ 
 \lt[ 1+\frac{a_{\rm fs}^q}{2} \rt] \lt[
  \cosh \frac{\dg_q\, t}{2} + \adg \sinh  \frac{\dg_q\, t}{2}
        \rt]  - \rt. \nn
&&\qquad\qquad\qquad \lt.
        \frac{a_{\rm fs}^q}{2} \lt[ 
                \adi \cos (\dm_q\, t) +  \ami \sin (\dm_q\, t) \rt] \rt\}
\label{untg}
\end{eqnarray}
with 
\begin{eqnarray}
\adi = \frac{1- \lt| \lambda_f \rt|^2}{1+ \lt| \lambda_f
        \rt|^2} , && \qquad
\ami = - \frac{2\, \imag \lambda_f}{1+ \lt| \lambda_f \rt|^2}
   \qquad\quad \mbox{and} \qquad
\adg = - \frac{2\, \real \lambda_f}{1+ \lt| \lambda_f   \rt|^2}
        \, . \label{defacp}
\end{eqnarray}
Hence one can obtain $a_{\rm fs}^q$ from the amplitude of the tiny
oscillations in \eq{untg}, once $\adi$ and $\ami$ are determined from the 
$\cos \dm_q\, t$ and $\sin \dm_q\, t$ terms of the time evolution in
the tagged $B_q (t)\to f$ decay. If $f$ is a CP eigenstate, $\adi$ and
$\ami$ are the direct and mixing-induced CP asymmetries. For example,
in $B_d \to J/\psi K_S$ one has  
$\lambda_f = -\exp (-2 i \beta)+{\cal O} (a_{\rm fs})$, so that one
can set $\adi=0$ and $\ami=-\sin(2\beta) $ in \eq{untg}. The
flavour-specific decays discussed in the previous section correspond
to the special case $\lambda_f=0$. 

\boldmath
\section{QCD corrections to $a_{\rm fs}^q$}
\begin{figure}
\vskip 0.2cm
\centerline{\psfig{figure=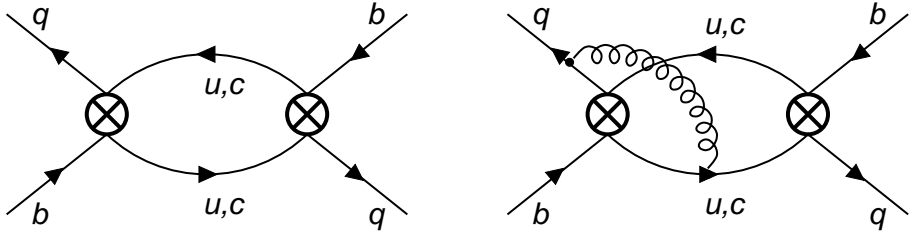,height=1.2in}}
\caption{Leading order contribution to $\Gamma_{12}$ (left) and a
sample NLO diagram (right). The crosses denote effective $\Delta B=1$
operators describing the $W$-mediated $b$ decay. 
The full set of NLO diagrams can
be found in \cite{bbgln1}.}\label{fig:lo}
\end{figure}

\unboldmath
$a_{\rm fs}^q=\imag \Gamma_{12}^q/M_{12}^q$ is proportional to two
powers of the charm mass $m_c$. A theoretical prediction in leading
order (LO) of QCD cannot control the renormalisation scheme of
$m_c$. Therefore the LO result $a_{\rm fs}^q$ suffers from a
theoretical uncertainty which is not only huge but also hard to
quantify. While next-to-leading order (NLO) QCD corrections to
$M_{12}^q$ are known for long \cite{bjw}, the computation of those to
$\Gamma_{12}^q$ has been completed only recently. The LO and a sample
NLO diagram are shown in Fig.~\ref{fig:lo}. 
The NLO result for the contribution with two identical up-type quark
lines (sufficient for the prediction of $\dg_s$) has been calculated
in \cite{bbgln1} and was confirmed in \cite{cflmt}. The contribution
with one up-quark and one charm-quark line was obtained recently in
\cite{cflmt} and \cite{bbln}. In order to compute $\Gamma_{12}^q$ one
exploits the fact that the mass $m_b$ of the $b$-quark is much larger
than the fundamental QCD scale $\lqcd$. The theoretical tool
used is the Heavy Quark Expansion (HQE), which yields a systematic
expansion of $\Gamma_{12}^q$ in the two parameters $\lqcd/m_b$ and
$\alpha_s(m_b)$ \cite{hqe}. $\Gamma_{12}^q$ and $M_{12}^q$ involve
hadronic ``bag'' parameters, which quantify the size of the
non-perturbative QCD binding effects and are difficult to compute.
The dependence on these hadronic parameters, however, largely cancels
from $a_{\rm fs}^q$.

\begin{figure}[t]
\vskip 0.2cm
\psfig{figure=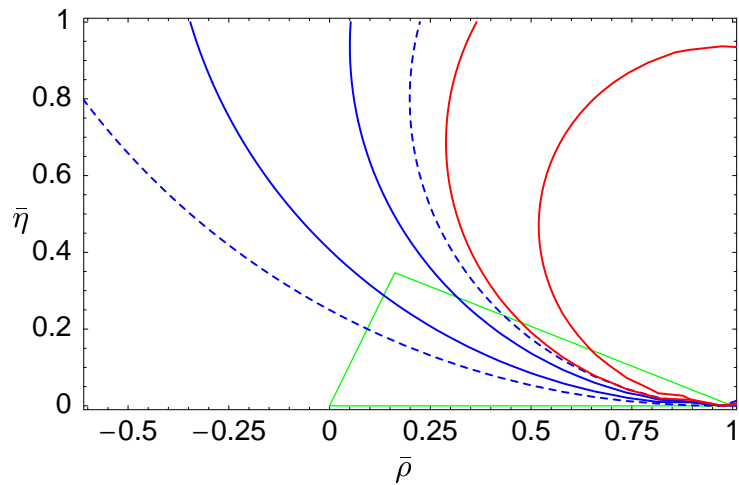,height=2.5in}
\caption{Constraint in the ($\bar\rho,\bar\eta$) plane
from $a^d_{\rm fs}$.
Area between solid pair of curves: NLO, for the cases 
$a^d_{\rm fs}=-5\times 10^{-4}$ (left) and $a^d_{\rm fs}=-10^{-3}$ (right).
Area between dashed curves: LO for $a^d_{\rm fs}=-5\times 10^{-4}$. 
The current best fit to the unitarity
triangle \cite{CKMWS} is also shown.}\label{fig:ckm}
\end{figure}

Including corrections of order $\alpha_s$ \cite{bbgln1,cflmt,bbln} and 
$\lqcd/m_b$ \cite{cflmt,BBD1,bbln} we predict \cite{bbln}
\begin{eqnarray}
a_{\rm fs}^d  &=& 10^{-4} \lt[ -\frac{\sin\beta}{R_t} ( 12.0\pm2.4  )
                + \lt( \frac{2\sin\beta}{R_t} -
                \frac{\sin 2 \beta}{R_t^2}  \rt) (0.2\pm 0.1)  \rt] .
\no
\end{eqnarray}
Here $\beta$ is the angle of the unitarity triangle measured in the CP
asymmetry of $B_d \to J/\psi K_S$. If $(\ov{\rho},\ov{\eta})$ denotes
the apex of the usual unitarity triangle, then
$R_t\equiv\sqrt{(1-\bar\rho)^2+\bar\eta^2}$ is the length of one of
its sides.  For the Standard Model fit to the unitarity triangle with
$\beta =22.4^\circ \pm 1.4^\circ$ and $R_t =0.91\pm 0.05$ \cite{CKMWS}
one finds:
\begin{eqnarray}
a_{\rm fs}^d &=& - (5.0 \pm 1.1)\cdot 10^{-4} \no
\end{eqnarray}
The impact of a future measurement of $a_{\rm fs}^d$ on the unitarity
triange is shown in Fig.~\ref{fig:ckm}.
The result for the $B_s$ system is
\begin{eqnarray}
a_{\rm fs}^s &=&  ( 12.0\pm2.4  ) \cdot 10^{-4} \, |V_{us}|^2 R_t \sin\beta
\; = \; (2.1\pm 0.4)\cdot 10^{-5} . \no
\end{eqnarray}

From \eq{dmdg} one finds that $\Delta\Gamma_q/\Delta M_q= -{\rm
Re}(\Gamma_{12}^q/M_{12}^q)$. This ratio was predicted to NLO in
\cite{bbgln1} for the $B_s$ system. With the new result of
\cite{cflmt,bbln} we can also predict $\Delta\Gamma_d/\Delta M_d$. Due
to a numerical accident, the Standard Model prediction for the ratio 
$\Delta\Gamma_q/\Delta M_q$ is essentially the same for $q=d$ and
$q=s$:
\begin{equation}\label{delgamd}
\frac{\Delta\Gamma_q}{\Delta M_q}=(4.0\pm 1.6)\times 10^{-3}\, ,
\qquad\qquad \frac{\Delta\Gamma_d}{\Gamma_d}=(3.0\pm 1.2)\times 10^{-3}.
\end{equation}
The precise values for the quark masses, ``bag'' factors and
$\alpha_s$ used for our numerical predictions can be found in Eq.~(7)
of \cite{bbln}.

We close our discussion with a remark about the $B_s$ system. It is
possible that new physics contributions render the \bbs\
oscillations so large that a measurement of $\dm_s$ will be
impossible. In general such new physics contribution will affect the
CP phase $\phi_s$ and suppress $\dg_s $ in
Eq.~(\ref{dmdg}). Different measurements of $\dg_s$ can then
determine $|\cos \phi_s|$ despite of the unobservably rapid \bbs\
oscillations \cite{g}. A measurement of the sign of $a_{\rm
fs}^s\propto \sin\phi_s$ (which will then be enhanced, unless $\dm_s$
is extreme) through e.g.\ \eq{aunt} will then reduce the four-fold
ambiguity in $\phi_s$ from the measurement of $|\cos \phi_s|$ to a
two-fold one.

\section*{Acknowledgements}
I thank the organisers for the invitation to this very pleasant and
stimulating Moriond conference. The presented  results stem from an
enjoyable collaboration with Martin Beneke, Gerhard Buchalla and 
Alexander Lenz \cite{bbln}. I am grateful to Guennadi Borisov for
pointing out a mistake in \eq{aunt}. 

Fermilab is operated by Universities Research Association Inc.\ under
Contract No.~DE-AC02-76CH03000 with the United States Department of
Energy.

\end{document}